# NoSQL Database: New Era of Databases for Big data Analytics - Classification, Characteristics and Comparison

A B M Moniruzzaman and Syed Akhter Hossain
*Department of Computer Science and Engineering*
*Daffodil International University*
*abm.mzkhan@gmail.com, aktarhossain@daffodilvarsity.edu.bd*

*Abstract*

*Digital world is growing very fast and become more complex in the volume (terabyte to petabyte), variety (structured and un-structured and hybrid), velocity (high speed in growth) in nature. This refers to as 'Big Data' that is a global phenomenon. This is typically considered to be a data collection that has grown so large it can't be effectively managed or exploited using conventional data management tools: e.g., classic relational database management systems (RDBMS) or conventional search engines. To handle this problem, traditional RDBMS are complemented by specifically designed a rich set of alternative DBMS; such as - NoSQL, NewSQL and Search-based systems. This paper motivation is to provide - classification, characteristics and evaluation of NoSQL databases in Big Data Analytics. This report is intended to help users, especially to the organizations to obtain an independent understanding of the strengths and weaknesses of various NoSQL database approaches to supporting applications that process huge volumes of data.*

## 1. Introduction

NoSQL, for "Not Only SQL," refers to an eclectic and increasingly familiar group of non-relational data management systems; where databases are not built primarily on tables, and generally do not use SQL for data manipulation [1]. NoSQL database management systems are useful when working with a huge quantity of data when the data's nature does not require a relational model.

NoSQL systems are distributed, non-relational databases designed for large-scale data storage and for massively-parallel data processing across a large number of commodity servers. They also use non-SQL languages and mechanisms to interact with data (though some new feature APIs that convert SQL queries to the system's native query language or tool). NoSQL database systems arose alongside major Internet companies, such as Google, Amazon, and Facebook; which had challenges in dealing with huge quantities of data with conventional RDBMS solutions could not cope [1]. They can support multiple activities, including exploratory and predictive analytics, ETL-style data transformation, and non mission-critical OLTP (for example, managing long-duration or inter-organization transactions). Originally motivated by Web 2.0 applications, these systems are designed to scale to thousands or millions of users doing updates as well as reads, in contrast to traditional DBMSs and data warehouses [13].

NewSQL systems are relational databases designed to provide ACID (Atomicity, Consistency, Isolation, Durability) -compliant, real-time OLTP (Online Transaction Processing) and conventional SQL-based OLAP in Big Data environments. These systems break through conventional RDBMS performance limits by employing NoSQL-style features such as column-oriented data storage and distributed architectures, or by employing technologies like in-memory processing [43], symmetric multiprocessing (SMP) [42] or Massively parallel Processing (MPP) [40].

## 2. Background

Of the many different data-models, the relational model has been dominating since the 80s, with implementations like Oracle databases [36], MySQL [35] and Microsoft SQL Servers [34] - also known as Relational Database Management System (RDBMS). Lately, however, in an increasing number of cases the use of relational databases leads to problems both because of deficits and problems in the modeling of data and constraints of horizontal scalability over several servers and big amounts of data. There are two trends that bringing these problems to the attention of the international software community:

1. The exponential growth of the volume of data generated by users, systems and sensors, further accelerated by the concentration of large part of this volume on big distributed systems like Amazon, Google and other cloud services.

2. The increasing interdependency and complexity of data accelerated by the Internet, Web2.0, social networks and open and standardized access to data sources from a large number of different systems.

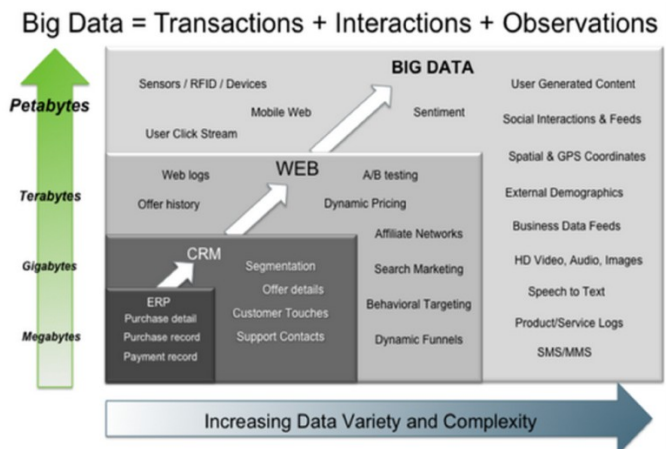

Figure 1: Big Data Transactions with Interactions and Observations. (Source: http://hortonworks.com/blog/7-key-drivers-for-the-big-data-market/ [41])

Organizations that collect large amounts of unstructured data are increasingly turning to non-relational databases, now frequently called NoSQL databases [4]. NoSQL databases focus on analytical processing of large scale datasets, offering increased scalability over commodity hardware [8]. Computational and storage requirements of applications such as for Big Data Analytics [9], Business Intelligence [10] and social networking over peta-byte datasets have pushed SQL-like centralized databases to their limits [5]. This led to the development of horizontally scalable, distributed non-relational data stores, called No-SQL databases, such as Google's Bigtable [6] and its open-source implementation HBase [33] and Facebook's Cassandra[7]. The emergence of distributed key-value stores, such as Cassandra and Voldemort [44], proves the efficiency and cost effectiveness of their approaches [3]. The main limitations with RDBMS are it is hard to scale with Data warehousing, Grid, Web 2.0 and Cloud applications [16]. Pokorny, J. (2011), focuses on NoSQL databases in context of cloud computing, particularly their horizontal scalability and concurrency model [17]. NoSQL databases are differing from Relational Database Management Systems (RDBMS) But NoSQL databases did not guarantee ACID properties [19].

The non-relational databases raised in recent years Motivated by requirements of Web 2.0 applications [2]. The strict relational schema can be a burden for web applications like blogs, which consist of many different kinds of attributes. Text, comments, pictures, videos, source code and other information have to be stored within multiple tables. Since such web applications are very agile, underlying databases have to be flexible as well in order to support easy schema evaluation [2]. Adding or removing a feature to a blog is not possible without system unavailability if a relational database is being used. NoSQL systems exhibit the ability to store and index arbitrarily big data sets while enabling a large amount of concurrent user requests [8].

## 3. Characteristics of NoSQL Databases

In order to guarantee the integrity of data, most of the classical database systems are based on transactions. This ensures consistency of data in all situations of data management. These transactional characteristics are also known as ACID (Atomicity, Consistency, Isolation, and Durability) [32]. However, scaling out of ACID-compliant systems has shown to be a problem. Conflicts are arising between the different aspects of high availability in distributed systems that are not fully solvable - known as the CAP- theorem [38]:
**Strong Consistency**: all clients see the same version of the data, even on updates to the dataset - e. g. by means of the two-phase commit protocol (XA transactions), and ACID,
**High Availability**: all clients can always find at least one copy of the requested data, even if some of the machines in a cluster is down,
**Partition-tolerance**: the total system keeps its characteristic even when being deployed on different servers, transparent to the client.
The CAP-Theorem postulates that only two of the three different aspects of scaling out are can be achieved fully at the same time. See figure 2.

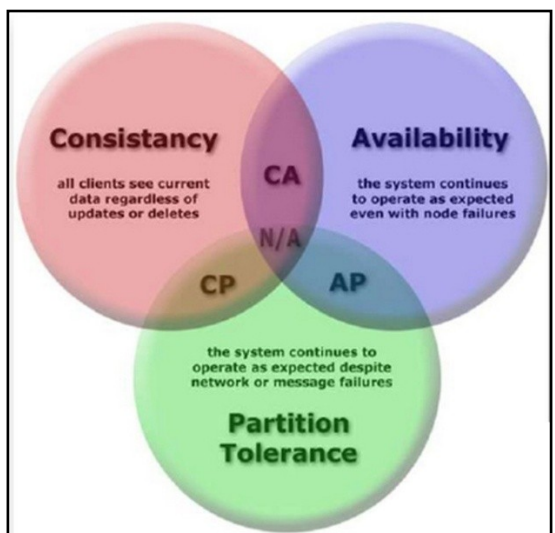

Figure 2: Characteristics of NoSQL Database (Source: nosqltips.blogspot.com )

Many of the NOSQL databases above all have loosened up the requirements on Consistency in order to achieve better Availability and Partitioning. This resulted in systems know as BASE (Basically Available, Soft-state, Eventually consistent) [39]. These have no transactions in the classical sense and introduce constraints on the data model to enable better partition schemes. Han, J., Haihong, E., Le, G., & Du, J. (2011) classifies NoSQL databases according to the CAP theorem [14]. Tudorica, B. G., & Bucur, C. (2011), compares using multiple criteria between several NoSQL databases [15].

Primary Uses of NoSQL Database (1) Large-scale data processing (parallel processing over distributed systems); (2) Embedded IR (basic machine-to-machine information look-up & retrieval); (3) Exploratory analytics on semi-structured data (expert level); (4) Large volume data storage (unstructured, semi-structured, small-packet structured).

Accordingly, they provide relatively inexpensive, highly scalable storage for high-volume, small-packet historical data like logs, call-data records, meter readings, and ticker snapshots (i.e., "big bit bucket" storage), and for unwieldy semi-structured or unstructured data (email archives, xml files, documents, etc.). Their distributed framework also makes them ideal for massive batch data processing (aggregating, filtering, sorting, algorithmic crunching (statistical or programmatic), etc.). They are good as well for machine-to-machine data retrieval and exchange, and for processing high-volume transactions, as long as ACID constraints can be relaxed, or at least enforced at the application level rather than within the DMS. Finally, these systems are very good exploratory analytics against semi-structured or hybrid data, though to tease out intelligence, the researcher usually must be a skilled statistician working in tandem with a skilled programmer.

## 4. Classification of NoSQL Databases

Leavitt, N. (2010), classifies NoSQL databases in three types: Key-value stores – e.g. SimpleDB [28]; column-oriented databases - e.g. Cassandra [29], HBase [33], Big Table [6]; and document-based stores - e.g. CouchDB [22], MongoDB [23]. In this section, we classify NoSQL Databases in four basic categories, each suited to different kinds of tasks –

(1) Key-Value stores; (2) Document databases (or stores); (3) Wide-Column (or Column-Family) stores; (4) Graph databases.

### 4.1 Key-Value stores

Typically, these DMS store items as alpha-numeric identifiers (keys) and associated values in simple, standalone tables (referred to as "hash tables"). The values may be simple text strings or more complex lists and sets. Data searches can usually only be performed against keys, not values, and are limited to exact matches. See figure: 3.

| Car | |
|---|---|
| Key | Attributes |
| 1 | Make: Nissan<br>Model: Pathfinder<br>Color: Green<br>Year: 2003 |
| 2 | Make: Nissan<br>Model: Pathfinder<br>Color: Blue<br>Color: Green<br>Year: 2005<br>Transmission: Auto |

Figure3: Key/Value Store NoSQL Database (Source: www.readwritewebcomimages.com)

Primary Use
The simplicity of Key-Value Stores makes them ideally suited to lightning-fast, highly-scalable retrieval of the values needed for application tasks like managing user profiles or sessions or retrieving product names. This is why Amazon makes extensive use of its own K-V system, Dynamo, in its shopping cart. Dynamo is a highly available key-value storage system that some of Amazon's core services use to provide highly available and scalable distributed data store [11].
Examples: Key-Value Stores- Dynamo (Amazon); Voldemort (LinkedIn); Redis; BerkeleyDB; Riak.

### 4.2 Document databases
Inspired by Lotus Notes, document databases were, as their name implies, designed to manage and store documents. These documents are encoded in a standard data exchange format such as XML, JSON (Javascript Option Notation) or BSON (Binary JSON). Unlike the simple key-value stores described above, the value column in document databases contains semi-structured data – specifically attribute name/value pairs. A single column can house hundreds of such attributes, and the number and type of attributes recorded can vary from row to row. Also, unlike simple key-value stores, both keys and values are fully searchable in document databases.

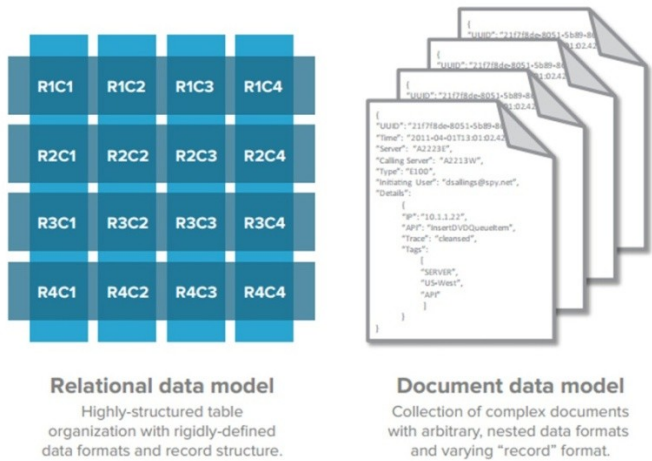

Figure 4: Document Store NoSQL Database (Source: http://gigaom.com/2011/07/29/couchbase-2-0-unql-sql-nosql/)

Primary Use
Document databases are good for storing and managing Big Data-size collections of literal documents, like text documents, email messages, and XML documents, as well as conceptual "documents" like de-normalized (aggregate) representations of a database entity such as a product or customer. They are also good for storing "sparse" data in general, that is to say irregular (semi-structured) data that would require an extensive use of "nulls" in an RDBMS (nulls being placeholders for missing or nonexistent values).Document Database Examples: CouchDB (JSON); MongoDB (BSON). MongoDB and CouchDB are open source and they are document oriented and schema free [12].

## 4.3 Wide-Column (or Column-Family) Stores (BigTable-implementations)

Like document databases, Wide-Column (or Column-Family) stores (hereafter WC/CF) employ a distributed, column-oriented data structure that accommodates multiple attributes per key. While some WC/CF stores have a Key-Value DNA (e.g., the Dynamo-inspired Cassandra), most are patterned after Google's Bigtable, the petabyte-scale internal distributed data storage system Google developed for its search index and other collections like Google Earth and Google Finance. These generally replicate not just Google's Bigtable data storage structure, but Google's distributed file system (GFS) and MapReduce parallel processing framework as well, as is the case with Hadoop, which comprises the Hadoop File System (HDFS, based on GFS) + Hbase (a Bigtable-style storage system) + MapReduce.

Primary Uses
This type of DMS is great for (1) Distributed data storage, especially versioned data because of WC/CF time-stamping functions. (2) Large-scale, batch-oriented data processing: sorting, parsing, conversion (e.g., conversions between hexadecimal, binary and decimal code values), algorithmic crunching, etc. (3) Exploratory and predictive analytics performed by expert statisticians and programmers. MapReduce is a batch processing method, which is why Google reduced the role of MapReduce in order to move closer to streaming/real-time index updates in Caffeine, its latest search infrastructure.

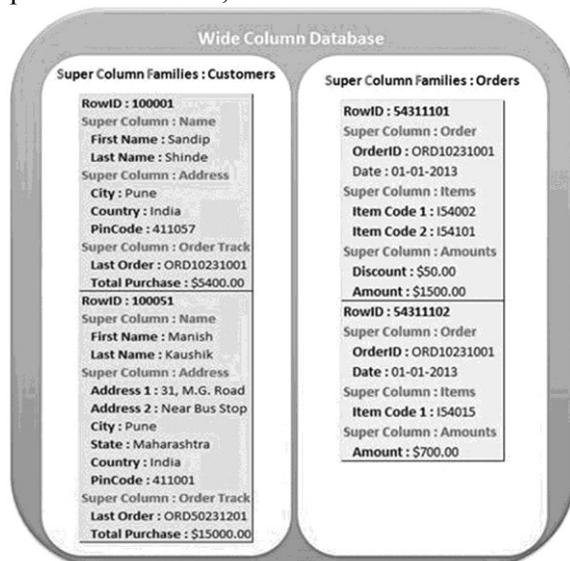

Figure 5: Wide-Column Store NoSQL Database
(Source: http://bi-bigdata.com/2013/01/13/what-is-wide-column-stores/)

Wide-Column/Column-Family Examples: Bigtable (Google); Hypertable; Cassandra (Facebook; used by Digg, Twitter); SimpleDB (Amazon); DynamoDB

### 4.4 Graph Databases

Graph databases replace relational tables with structured relational graphs of interconnected key-value pairings. They are similar to object-oriented databases as the graphs are represented as an object-oriented network of nodes (conceptual objects), node relationships ("edges") and properties (object attributes expressed as key-value pairs). They are the only of the four NoSQL types discussed here that concern themselves with relations, and their focus on visual representation of information makes them more human-friendly than other NoSQL DMS.

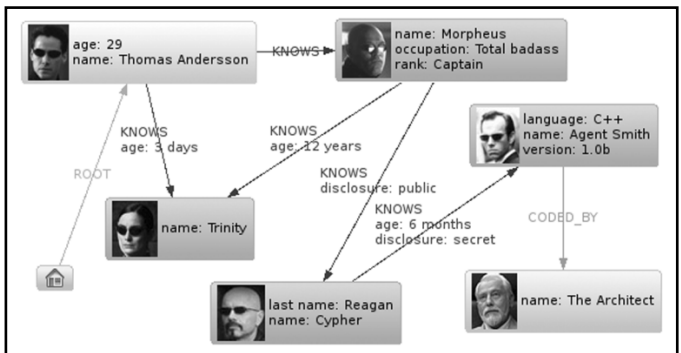

Figure 6: Graph NoSQL Database (Source: http://blog.neo4j.org/2010/02/top-10-ways-to-get-to-know-neo4j.html )

Primary uses

In general, graph databases are useful when you are more interested in relationships between data than in the data itself: for example, in representing and traversing social networks, generating recommendations (e.g., upsell or cross-sell suggestions), or conducting forensic investigations (e.g., patterndetection). Note these DMS are optimized for relationship "traversing," not for querying. If you want to explore relationships as well as querying and analyzing the values embedded within them (and/or to be able to use natural language queries to analyze relationships), then a search-based DMS is a better choice.

Graph Database Examples: Neo4j; InfoGrid; Sones GraphDB; AllegroGraph; InfiniteGraph

NoSQL, in its incarnation at least, is a relatively new technology. However, it has already attracted a significant amount of attention due to its use by massive websites like Amazon, Yahoo, Facebook, which have data utilization rates that bring relational databases to a crawl.

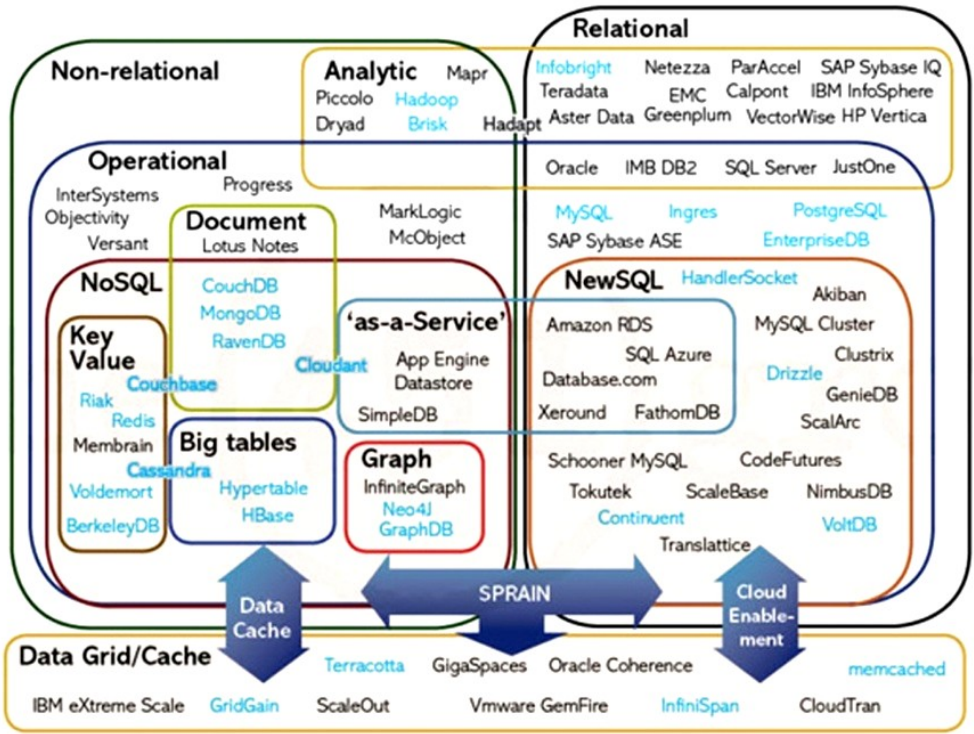

Figure 7: Current State of NoSQL Databases (Source: techielicous.com/2011/11/02/nosql-in-the-real-world )

Dozens of products self-identify as NoSQL, and each has its own unique architecture and design. Even the data storage paradigm varies among implementations. Columnar, key-value and document oriented repositories exist. NoSQL began within the domain of open source and a few small vendors, but continued growth in data and NoSQL has enticed many new players into the market. See figure: 7. NoSQL solutions are attractive because they can handle huge quantities of data, relatively quickly, across a cluster of commodity servers that share resources. In additon, most NoSQL solutions are open source, which gives them a price advantage over conventional commercial databases.

## 5. Comparison of NoSQL Database

In this section, we provide evaluation some of NoSQL databases (four categories) with a matrix on basis of few attributes- design, integrity, indexing, distribution, system. See table 1.

| Attributes | | NoSQL Databases | | | | | | | | |
|---|---|---|---|---|---|---|---|---|---|---|
| Database model | | Document-Stored | | Wide-Column Stored | | | | Key-Value Stored | | Graph-oriented |
| | Features | MongoDB | CouchDB | DynamoBD | HBase | Cassandra | Accumulo | Redis | Riak | Neo4j |
| **Design & Features** | Data storage | Volatile memory File System | Volatile memory File System | SSD | HDFS | | Hadoop | Volatile memory File System | Bitcask LevelDB Volatile memory | File System Volatile memory |
| | Query language | Volatile memory File System | JavaScript Memcached-protocol | API calls | API calls REST XML Thrift | API calls CQL Thrift | | API calls | HTTP JavaScript REST Erlang | API calls REST SparQL Cypher Tinkerpop Gremlin |
| | Protocol | Custom, binary (BSON) | HTTP, REST | - | HTTP/REST Thrift | Thrift & custom binary CQL3 | Thrift | Telnet-like | HTTP, REST | HTTP/REST embedding in Java |
| | Conditional entry updates | Yes | Yes | Yes | Yes | No | Yes | No | No | |
| | MapReduce | Yes | Yes | Yes | Yes | Yes | Yes | No | Yes | No |
| | Unicode | Yes | Yes | Yes | Yes | Yes | Yes | Yes | Yes | Yes |
| | TTL for Entries | Yes | Yes | No | Yes | Yes | Yes | Yes | Yes | |
| | Compression | Yes | Yes | - | Yes | Yes | Yes | Yes | Yes | |
| **Integrity** | Integrity model | BASE | MVCC | ASID | Log Replication | BASE | MVCC | - | BASE | ASID |
| | Atomicity | Conditional | Yes | Yes | Yes | Yes | Conditional | Yes | No | Yes |
| | Consistency | Yes | Yes | Yes | Yes | Yes | Yes | Yes | No | Yes |
| | Isolation | No | Yes | Yes | No | No | - | Yes | Yes | Yes |
| | Durability (data storage) | Yes | Yes | Yes | Yes | Yes | Yes | Yes | - | Yes |
| | Transactions | No | No | No | Yes | No | Yes | Yes | No | Yes |
| | Referential integrity | No | No | No | No | No | No | Yes | No | Yes |
| | Revision control | No | Yes | Yes | Yes | No | Yes | No | Yes | No |
| **Indexing** | Secondary Indexes | Yes | Yes | No | Yes | Yes | Yes | - | Yes | - |
| | Composite keys | Yes | Yes | Yes | Yes | Yes | Yes | - | Yes | |
| | Full text search | No | No | No | No | No | Yes | No | Yes | Yes |
| | Geospatial Indexes | Yes | No | No | No | No | Yes | - | - | Yes |
| | Graph support | No | No | No | No | No | No | No | Yes | Yes |
| **Distribution** | Horizontal scalable | Yes | Yes | Yes | Yes | Yes | Yes | | Yes | No |
| | Replication | Yes | Yes | Yes | Yes | Yes | Yes | Yes | Yes | Yes |
| | Replication mode | Master-Slave-Replica Replication | Master-Slave Replication | - | Master-Slave Replication | Master-Slave Replication | - | Master-Slave Replication | Multi-master replication | - |
| | Sharding | Yes | Yes | Yes | Yes | Yes | Yes | No | Yes | Yes |
| | Shared nothing architecture | Yes | Yes | Yes | Yes | Yes | - | - | Yes | - |
| **System** | Value size max. | 16MB | 20MB | 64KB | 2TB | 2GB | 1EB | - | 64MB | |
| | Operating system | Cross-platform | Ubuntu Red Hat Windows Mac OS X | Cross-platform | Cross-platform | Cross-platform | NIX 32 entries Operating system | Linux *NIX Mac OS X Windows | Cross-platform | Cross-platform |
| | Programming language | C++ | Erlang C++ C Python | Java | Java | Java | Java | C C++ | Erlang | Java |

Table 1: Comparison some of NoSQL databases (four categories) with a matrix on basis of few attributes- design, integrity, indexing, distribution, system.

## 6. Adoption of NoSQL Database

The acronym NoSQL was coined in 1998. Many people think NoSQL is a derogatory term created to poke at SQL. In reality, the term means Not Only SQL. The idea is that both technologies can coexist and each has its place. The NoSQL movement has been in the news in the past few years as many of the Web 2.0 leaders have adopted a NoSQL technology. Companies like Facebook, Twitter, Digg, Amazon, LinkedIn and Google all use NoSQL in one way or another.

Couchbase Survey [37] was conducted in the year 2012. Key data points from the Couchbase NoSQL survey include:
- Nearly half of the more than 1,300 respondents indicated they have funded NoSQL projects in the first half of this year. In companies with more than 250 developers, nearly 70% will fund NoSQL projects over the course of 2012.
- 49% cited rigid schemas as the primary driver for their migration from relational to NoSQL database technology. Lack of scalability and high latency/low performance also ranked highly among the reasons given for migrating to NoSQL (see chart below for more details).
- 40% overall say that NoSQL is very important or critical to their daily operations, with another 37% indicating it is becoming more important.

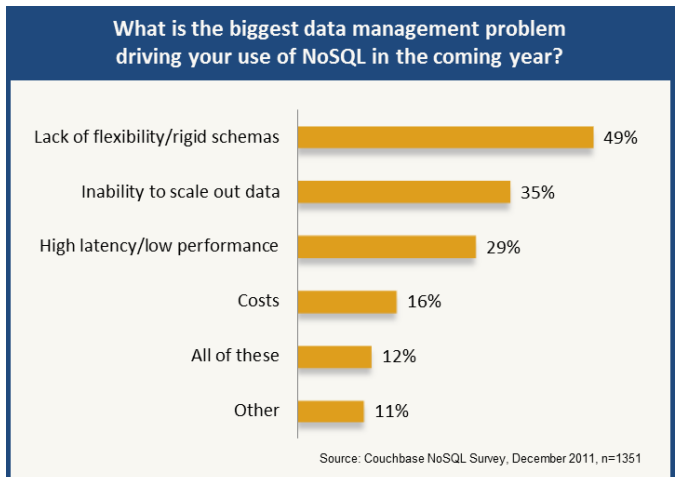

Figure 8: Key problems-driving to NoSQL databases (source: [37])

Organizations that have massive data storage needs are looking seriously at NoSQL. And NoSQL Database expert are highly demanded for most of the developing organizations. This graph shows job trends of five NoSQL Databases from Indeed.com:

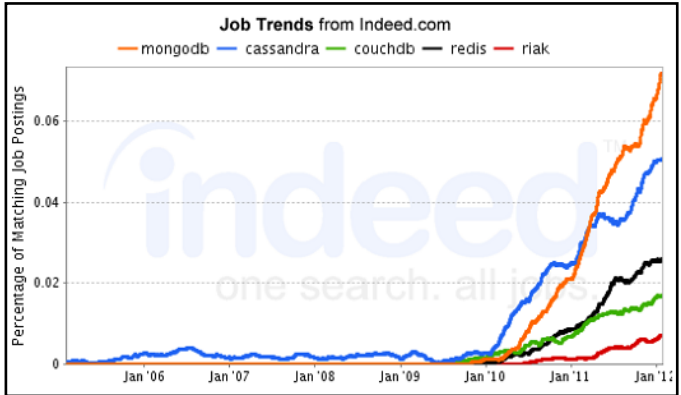

Figure 9: job trends of five NoSQL Databases (source: Indeed.com)

MongoDB's growth means that it has cemented its place as the most popular NoSQL database, according to LinkedIn profile mentions. As the chart below illustrates, it now accounts for 45% of all mentions of NoSQL technologies in LinkedIn profiles. See figure 0.

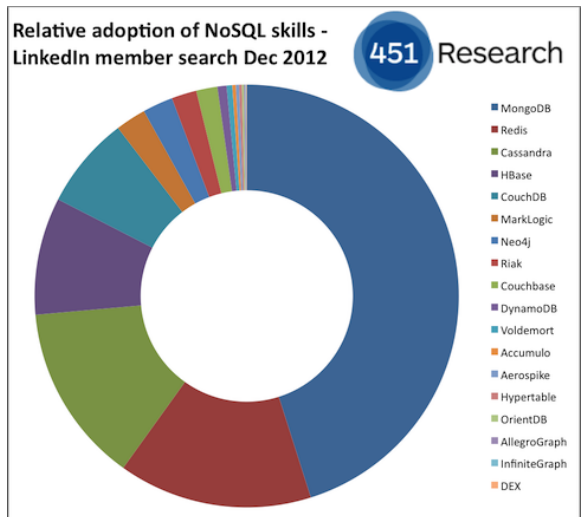

Figure 10: NoSQL LinkedIn Skills Index – December 2012 (source: http://blogs.the451group.com)

Organizations that collect large amounts of unstructured data are increasingly turning to non-relational databases. NoSQL databases focus on analytical processing of large scale datasets, offering increased scalability over commodity hardware. NoSQL systems exhibit the ability to store and index arbitrarily big data sets while enabling a large amount of concurrent user requests.

## 7. Conclusion

Computational and storage requirements of applications such as for Big Data Analytics, Business Intelligence and social networking over peta-byte datasets have pushed sql-like centralized databases to their limits [8]. This led to the development of horizontally scalable, distributed non-relational No-SQL databases. We speculate some of the major (primarily) uses of NoSQL Databeses: Large-scale data processing (parallel processing over distributed systems); Embedded IR (basic machine-to-machine information look-up & retrieval); Exploratory analytics on semi-structured data (expert level); Large volume data storage (unstructured, semi-structured, small-packet structured)

NoSQL is a large and expanding field, for the purposes of this paper - characteristics (features and benefits of NoSQL databases); classification (categories four on their features); comparison and evaluation (with a matrix on basis of few attributes- design, integrity, indexing, distribution, system) of different types of NoSQL databases; and current state of adoption of NoSQL databases. This study report motivation to provide an independent understanding of the strengths and weaknesses of various NoSQL database approaches to supporting applications that process huge volumes of data; as well as to provide a global overview of this non-relational NoSQL databases.

# Authors

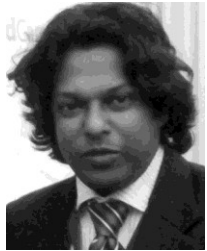

**A B M Moniruzzaman** Received his B.Sc (Hon's) degree in Computing and Information System (CIS) from London Metropolitan University, London, UK and M.Sc degree in Computer Science and Engineering (CSE) from Daffodil International University, Dhaka, Bangladesh in 2005 and 2013, respectively. Currently he is working on research on Cloud Computing and Big Data Analytics as a research associate at RCST (Research Center for Science and Technology) at Daffodil International University (DIU), Dhaka, Bangladesh. Besides, his voluntarily works as reviewer of few international journals including IEEE, Elsevier. His research interests include Cloud Computing, Cloud Applications, Open-source Cloud, Cloud Management Platforms, Building Private and Hybrid Cloud with FOSS software, Big Data Management, Agile Software Development, Hadoop, MapReduce, Parallel and Distributed Computing, Clustering.

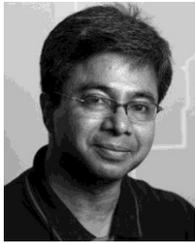

**Prof. Dr. Syed Akhter Hossain** is Post Doctoral Fellow, Informatics and Systems Engineering, LIESP Laboratory, Universite Lyon2, Lyon, France. He received PhD in Computer Science and Engineering from University of Dhaka, Bangladesh and MSc degree in Applied Physics and Electronics, First Class (First), and BSc (Hons) in Applied Physics and Electronics, First Class (First), Gold Medalist from Rajshahi University, Rajshahi, Bangladesh. Currently he is working as Professor and Head, Department of Computer Science and Engineering, Daffodil International University, Dhaka, Bangladesh. Besides, he received best professor award from Singapore and has got more than 60 international publications including journals and proceedings and 3 book chapters with IGI Global and John Wiley. He is a member of ACM, and member of IEEE. His research areas includes simulation and modeling distributed system design and implementation, signal and image processing, internet and web engineering, network planning and management, database and data warehouse modeling.